\documentclass{aa}  

\usepackage{graphicx}
\usepackage{txfonts}
\newcommand{\name}{HZ4}
\newcommand{\cii}{\mbox{[\ion{C}{ii}]}}
\newcommand{\hi}{\mbox{\ion{H}{i}}}
\newcommand{\oi}{\mbox{[\ion{O}{i}]}}
\newcommand{\oiii}{\mbox{[\ion{O}{iii}]}}
\newcommand{\sfir}{$\Sigma_{\rm FIR}$}
\newcommand{\kcii}{$\kappa_{\rm [CII]}$}
\newcommand{\nh}{$n_{\rm H}$}

\begin{document}

   \title{A kiloparsec view of a typical star-forming galaxy when the Universe was $\sim1$~Gyr old}

   \subtitle{I. Outflow, halo, and interstellar medium properties}

  \author{R. Herrera-Camus\inst{1},
          N. F{\"o}rster Schreiber\inst{2}, 
          R. Genzel\inst{2},
          L. Tacconi\inst{2},
          A. Bolatto\inst{3},
          R. L. Davies\inst{2},
          D. Fisher\inst{4,5},
          D. Lutz\inst{2},
          T. Naab\inst{6},
          T. Shimizu\inst{2},
          K. Tadaki\inst{7}
          \and
          H. {\"U}bler\inst{2}
	 }
	\authorrunning{Herrera-Camus et al.}

   \institute{Departamento de Astronom\'ia, Universidad de Concepci\'on, Barrio 			Universitario, Concepci\'on, Chile\\
              \email{rhc@astro-udec.cl}
              \label{1}
	\and
          Max-Planck-Institut f\"ur extraterrestische Physik (MPE), Giessenbachstr., D-85748 Garching, Germany
	\label{2}
	\and
	Department of Astronomy, University of Maryland, 
	College Park, MD 20742, USA 
	\label{3}
	\and
	Centre for Astrophysics and Supercomputing, Swinburne Univ. of Technology, P.O. Box 218, Hawthorn, VIC 3122, Australia
	\label{4}
	\and
	ARC Centre of Excellence for All Sky Astrophysics in 3 Dimensions (ASTRO 3D), Australia
	\label{5}
	\and
	Max-Planck Institute for Astrophysics, Karl Schwarzschildstrasse 1, D-85748 Garching, Germany
	\label{6}
	\and
	 National Astronomical Observatory of Japan, 2-21-1 Osawa, Mitaka, Tokyo 181-8588, Japan
	 \label{7}
             }

   \date{Received ; accepted }

% \abstract{}{}{}{}{} 
% 5 {} token are mandatory
 
  \abstract
{We present new Atacama Large Millimeter/Submillimeter Array (ALMA) observations of the \cii~158~$\mu$m transition and the dust continuum in \name, a typical star-forming galaxy when the Universe was only $\sim1$~Gyr old ($z\approx5.5$). Our high $\approx0.3$\arcsec\ spatial resolution allow us to study the relationships between \cii\ line emission, star formation rate (SFR), and far-infrared (FIR) emission on spatial scales of $\sim2$~kpc. In the central $\sim$4~kpc of \name\, the \cii/FIR ratio is $\sim3\times10^{-3}$ on global scales as well as on spatially-resolved scales of $\sim$2 kpc, comparable to the ratio observed in local moderate starburst galaxies such as M82 or M83. For the first time in an individual normal galaxy at this redshift, we find evidence for outflowing gas from the central star-forming region in the direction of the minor-axis of the galaxy. The projected velocity of the outflow is $\sim400$~km~s$^{-1}$, and the neutral gas mass outflow rate is $\sim3-6$ times higher than the SFR in the central region. Finally, we detect a diffuse component of \cii\ emission, or \cii-halo, that extends beyond the star-forming disk and has a size of $\sim12$~kpc in diameter. Most likely the outflow, which has a velocity approximately half the escape velocity of the system, is in part responsible for fueling the \cii\ extended emission. Together with the kinematic analysis of \name\ (presented in a forthcoming paper), the analysis supports that \name\ is a typical star-forming disk at $z\sim5$ with interstellar medium (ISM) conditions similar to present-day galaxies forming stars at a similar level, driving a galactic outflow that may already play a role in its evolution.}


   \maketitle
%
%-------------------------------------------------------------------

\section{Introduction}

  The complex interplay between star formation, stellar feedback, and accretion from the circumgalactic medium (CGM), known as the baryon cycle, is responsible for driving galaxy growth and evolution \citep[e.g.,][]{rhc_tumlinson17}. Understanding these competing processes requires observations of the stars and multiple phases of the gas across cosmic time. Here, the Atacama Large Millimeter/submillimeter Array (ALMA) has opened a window to explore the cold neutral and molecular gas in early galaxies at an unprecedented level of detail. 

This paper focuses on the study of a typical galaxy at $z\approx5.5$, when the Universe was only $\sim1$ billion years old. Observations of the cold gas based on CO lines for normal galaxies at this redshift are challenging \citep[e.g.,][]{rhc_pavesi19,rhc_neeleman20}. A powerful alternative tracer is the \cii~157.74~$\mu$m fine-structure transition, which mostly traces the cold, neutral gas with a minor contribution from the ionized phase \citep[e.g., ][]{rhc_croxall17,rhc_diaz-santos17,rhc_cormier19}. The \cii\ line is tightly connected to the star formation activity, because the far-ultraviolet (FUV) photons produced in star-forming regions heat the gas via the photoelectric effect on dust grains, which then cools mainly via the \cii\ transition \citep[e.g.,][]{rhc_wolfire95,rhc_delooze14,rhc_rhc15,rhc_schaerer20}. In that sense, the comparison between tracers of the gas heating --such as the star formation rate (SFR) or the far-infrared luminosity (FIR)-- and the \cii\ line luminosity, can help to constrain the heating-cooling balance in the interstellar medium (ISM).

\begin{figure*}
\centering
\includegraphics{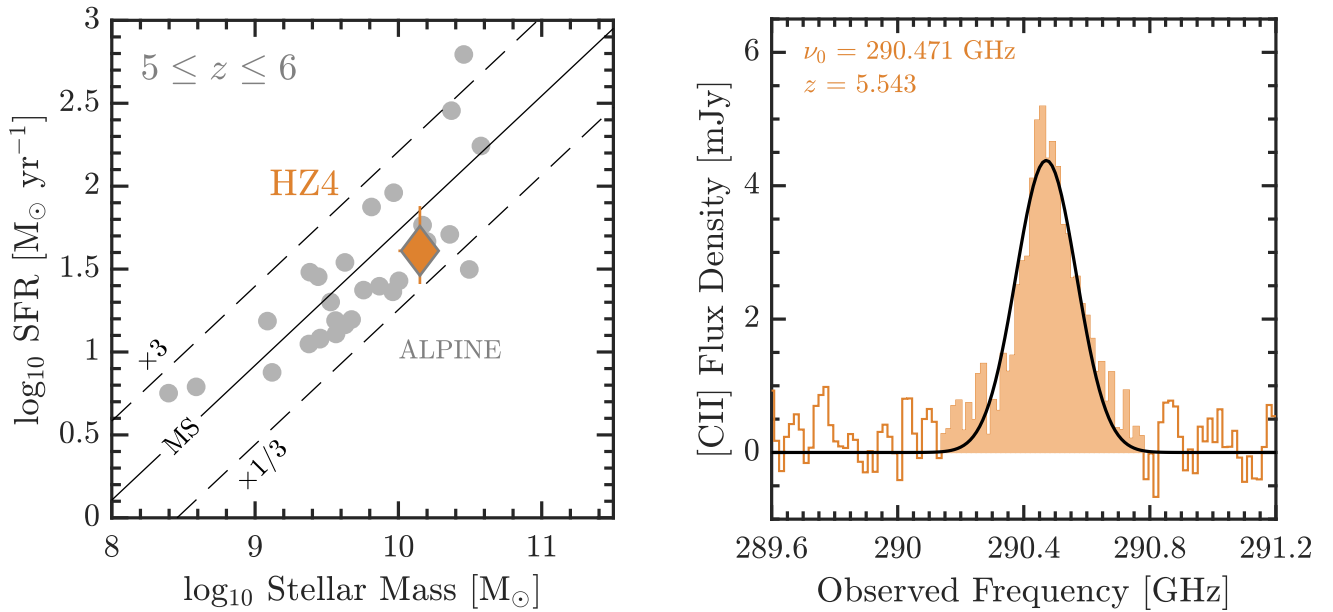}
\caption{({\it Left}) Star formation rate - stellar mass plane between $5\leq z \leq 6$ showing the position of the main-sequence (MS) of star-forming galaxies \citep[black line;][]{rhc_speagle14} and the $z\sim5$ galaxies with \cii\ detections in the ALPINE survey \citep[gray circles; ][]{rhc_lefevre19,rhc_faisst20}. \name, shown as an orange diamond, lies only $\sim0.2$~dex above the main-sequence and can be considered a {\it typical} star-forming galaxy at this redshift \citep{rhc_faisst20}. ({\it Right}) Continuum subtracted \cii~158~$\mu$m spectrum of \name\ extracted inside a circular aperture of 0.9\arcsec\  radius ($\sim5.4$~kpc). The best Gaussian fit is centered at an observed frequency of $\nu_0=290.471$~GHz, which corresponds to a redshift of $z=5.54$.
\label{fig:ciispec}
}
\end{figure*}

Recent ALMA observations of main-sequence galaxies at $z\sim4-5$ have revealed different aspects of the baryon cycle in action in these systems. On the one hand, stacking of \cii\ spectra reveals signatures of outflows with velocities of $\sim500$~km~s$^{-1}$, and global mass outflow rates comparable to the SFRs of their hosts \citep{rhc_gallerani18,rhc_ginolfi20}. At similar redshift, \cite{rhc_sugahara19} also detect neutral gas outflows with median velocities of $\sim400$~km~s$^{-1}$ from the analysis of the stacked spectra of metal-absorption lines. As simulations suggest \citep[e.g., ][]{rhc_pizzati20}, these outflows are most likely connected to the observed presence of a diffuse component of \cii\ emission around galaxies. These \cii-halos extend for scales of $\sim5-10$~kpc beyond the extent of the star-forming disk traced by the rest-frame UV/optical emission, and can be considered part of the CGM \citep[][]{rhc_fujimoto19,rhc_fujimoto20,rhc_ginolfi20}. 

The results discussed above are based on observations of large sample of galaxies at $4 \lesssim z \lesssim 6$ from the ALMA Large Program to Investigate C+ at Early Times \citep[ALPINE; ][]{rhc_lefevre19}, and the earlier work by \cite{rhc_capak15}. The spatial resolution achieved by these programs was modest, typically $\sim5-9$~kpc \cite[for reference, the median UV effective radius of $z\sim5$ galaxies is $\sim$1~kpc; e.g., ][]{rhc_shibuya15}. To date, the majority of the \cii\ observations of early galaxies that achieved kiloparsec scale resolution (or better) targeted bright, submillimeter galaxies \citep[e.g., ][]{rhc_debreuck14,rhc_neri14,rhc_diaz-santos16,rhc_gullberg18,rhc_litke19,rhc_tadaki20}. The few cases that focused on normal (i.e., main-sequence) galaxies include the discovery of a regular, rotating disk in a neutral \hi\ absorber \citep{rhc_neeleman20}, a system with a complex kinematic structure that could be the result of a triple merger \citep{rhc_riechers14}, and two Lyman break galaxies at $z\sim7$ that show disk-like velocity gradients \citep{rhc_smit18}.

Although progress is being made in the study of main-sequence galaxies at $z\sim5$, we are still lacking a detailed characterization of their ISM, the detection of outflow signatures in individual galaxies, and a deeper understanding of the impact of outflows on the star formation activity and the existence and maintenance of the \cii-halos. Motivated by these open issues,  in this paper we present new \cii\ and dust continuum ALMA observations of HZ4, a typical star-forming galaxy at $z=5.5$. The spatial resolution and sensitivity achieved by our observations (see Section~\ref{sec:data} and \ref{sec:results}) allow us to study the ISM (Section~\ref{Sec:cii2fir}), outflow (Section~\ref{Sec: wind}), and \cii\ extended emission (Section~\ref{Sec:halo}) properties at a kiloparsec scale level. The analysis of the kinematic properties of \name\ will be discussed in a forthcoming paper (Paper~II; Herrera-Camus et al. in prep.). Throughout this paper, we adopt a cosmology with $H_{0} = 67.4$~km~s$^{-1}$~Mpc$^{-1}$ and $\Omega_{\rm M}=0.315$ \citep{rhc_planck18}, which results in a scale of $6.09$~kpc/\arcsec\ for a source at $z=5.54$.

%--------------------------------------------------------------------

\begin{figure*}
\centering
\includegraphics{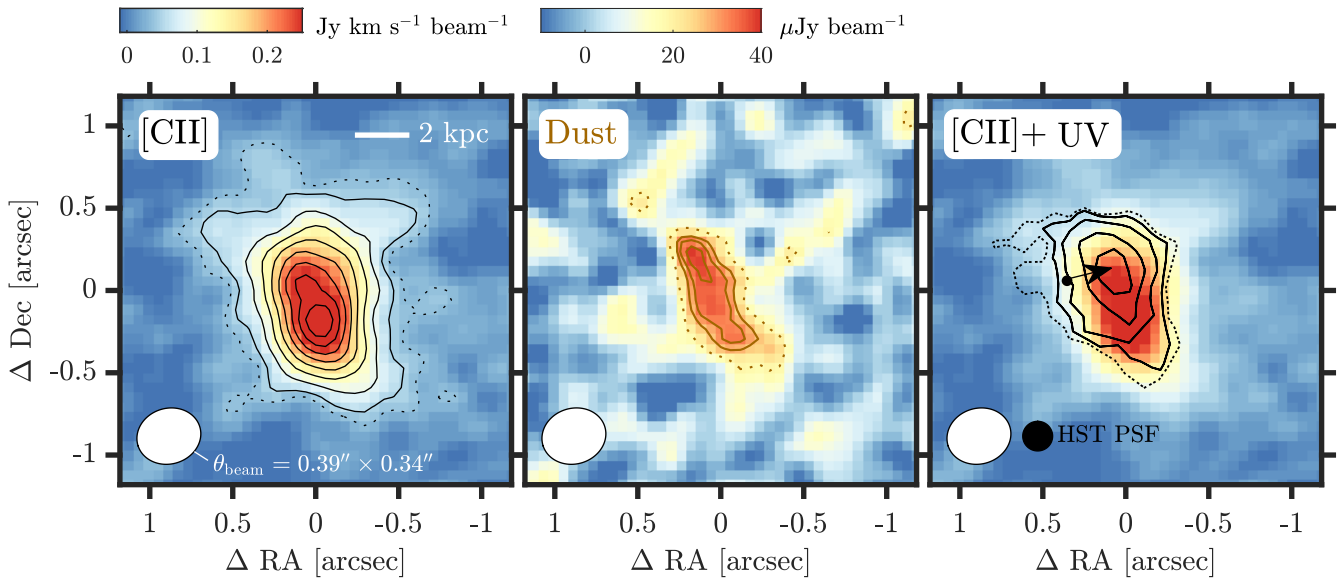}
\caption{{(\it Left)} \cii~158~$\mu$m integrated intensity of \name. The contours start at 2$\sigma$ (dotted line) and then increase from 3$\sigma$ to 17$\sigma$ in steps of two (solid lines). The ALMA synthesized beam ($\theta=0.39\arcsec \times 0.34\arcsec$) is shown in the bottom-left corner. {(\it Center)}  Dust continuum emission at rest-frame 160~$\mu$m. The contours level are 2 (dotted brown line), and 3, 3.5 and 4$\sigma$ (solid brown lines). {(\it Right)} Contours of rest-frame UV emission as observed by HST/WFC3 F160W \citep{rhc_barisic17} on top of the \cii\ integrated intensity map. The contour levels are 2 (dashed black line), 3, 5, 10 and 20$\sigma$ (solid black lines). The contours are shifted in the direction of the black arrow from the original position marked by the black dot. The HST WFC3 F160W point-spread function (PSF) is shown in black in the bottom left corner next to the ALMA beam in white.\label{fig_mom}
}
\end{figure*}

\section{Observations and data reduction} \label{sec:data}

Between November 2018 and April 2019 (Cycle~6), \name\ (R.A.~9:58:28.5, Dec.~$+$2:03:06.7) was observed for 8.4 hr total (4.7~hr on-source) using ALMA as part of project 2018.1.01605.S (PI Herrera-Camus). The observations were carried out in the C43-4 configuration. We centered one spectral window around the \cii\ line ($\nu_{\rm rest}=1900.537$~GHz), which for the source is redshifted to $\nu_{\rm obs}=290.386$~GHz and falls in Band~7. The remaining three spectral windows were used to detect the dust continuum at rest-frame 160~$\mu$m, near the peak of the FIR spectral energy distribution in star-forming galaxies. The quasar J1058+0133 was used as the flux and bandpass calibrator, and the quasar J0948+0022 was used as the phase calibrator.

The data were processed using the Common Astronomy Software Applications package \citep[CASA;][]{rhc_casa} version 5.6.2. The cubes were cleaned using the CASA/\texttt{tclean} task with Briggs weighting (robust=+0.5) and down to the 1$\sigma$ level using the ``auto-multithresh" automasking algorithm \citep{rhc_kepley20}. We also adopted the Multi-Scale Clean algorithm \citep{rhc_cornwell08} to image both structures that are compact and unresolved as well as extended. The resulting synthesized beam size was $\theta_{\rm beam}=0.39\arcsec\times0.34\arcsec$ ($2.3\times2.0$~kpc) at a position angle of ${\rm PA}=-72.4^{\rm o}$. The maximum recoverable scale is 3.26\arcsec,\footnote{According to Table~7.1 in the Cycle 6 ALMA Technical Handbook.} which is larger than the angular scale of the \cii\ emission structures detected in our observations (see Section~\ref{Sec:halo}). The rms noise for the line cube is 0.15~mJy~beam$^{-1}$ in 16~km~s$^{-1}$ channels, and 8.5~$\mu$Jy~beam$^{-1}$ in the continuum maps. All maps were primary beam-corrected.

\section{Results}\label{sec:results}

\subsection{\cii\ line and dust continuum emission in the main-sequence star-forming galaxy \name} \label{Sec:emission} 

\name\ is a Lyman break selected galaxy in the COSMOS field \citep{rhc_koekemoer07} with a \cii-based spectroscopic redshift of $z=5.54$ \citep{rhc_capak15,rhc_hasinger18,rhc_bethermin20}. As Figure~\ref{fig:ciispec} (left panel) shows, with a stellar mass of log$_{10}(M_{\star}/M_{\odot})=10.15^{+0.13}_{-15}$ 
%log$_{10}(M_{\star}/M_{\odot})=9.67\pm0.21$ 
and a star formation rate of ${\rm SFR=40.7^{+35}_{-15}}~M_{\odot}~{\rm yr}^{-1}$ \citep{rhc_faisst20},
%${\rm SFR(UV+IR)=54^{+51}_{-18}}~M_{\odot}~{\rm yr}^{-1}$ \citep{rhc_capak15},
 \name\ lies on the main-sequence of star-forming galaxies at $z\sim5$ \citep[e.g.,][]{rhc_speagle14}. The gray circles show galaxies between $5\leq z \leq6$ detected in \cii\ line emission as part of the ALMA ALPINE Large Program \citep{rhc_lefevre19,rhc_bethermin20}. In contrast to our deep, high-resolution observation, ALPINE galaxies have typical on-source integration times of only $t\sim15-25$~min with angular resolution in the range $\theta_{\rm beam}\sim0.8-1.5$\arcsec.

Figure~\ref{fig:ciispec} (right panel) shows the global \cii\ spectrum of \name\ extracted within a circular aperture of 0.9\arcsec ($\sim5.4$~kpc) radius centered at the peak of the \cii\ emission (RA 09:58:28.5, Dec. +02:03:06.6). The line is detected with a global signal-to-noise ratio of ${\rm S/N}\approx16$. The best single Gaussian fit is centered at an observed frequency of $\nu_0=290.471\pm2\times10^{-3}$~GHz, which corresponds to a redshift of $z=5.543$, consistent with previous measurements. The peak flux density and linewidth ( full width at half maximum; FWHM) are $4.79\pm0.22$~mJy and $232\pm8$~km~s$^{-1}$, respectively. The total flux measured is $S_{\rm [CII]}=1.12\pm0.07$~Jy~km~s$^{-1}$. This is similar to the flux measured by \cite{rhc_capak15} of $S_{\rm [CII]}=1.14$~Jy~km~s$^{-1}$ based on observations with an angular resolution of $\theta_{\rm beam}=0.90\arcsec\times0.48\arcsec$, and about 20\% larger than the total flux measured based on observations from the ALPINE survey \citep[under the name DEIMOS\_COSMOS\_494057;][]{rhc_lefevre19,rhc_bethermin20} with an angular resolution of $\theta_{\rm beam}=1.01\arcsec\times0.85\arcsec$. Most likely due to the high S/N of our observations, we do not observe a lower \cii\ global flux in our high angular resolution data as reported in other high-redshift galaxies due to the effect of surface brightness dimming discussed by \cite{rhc_carniani20}.

% 1.01"0.85"
%, similar to the measurement by \cite{rhc_capak15} from lower angular resolution observations ($\theta_{\rm beam}=0.90\arcsec\times0.48\arcsec$). % COMPARISON IN THE CONTEXT OF CARNIANI

The left panel of Figure~\ref{fig_mom} shows the \cii\ line intensity map integrated in the velocity range between $-300$ and $+300$~km~s$^{-1}$, which corresponds to the region in the spectrum of Figure~\ref{fig:ciispec} filled in orange. The galaxy morphology resembles an extended disk that looks slightly warped in the southwest region. A two-dimensional Gaussian fitting yields a deconvolved size (at FWHM) of $0.80\arcsec \times 0.47\arcsec$ ($\pm0.05\arcsec$) ($4.9\times2.9$~kpc) with the major axis at a PA of $17.7^{\rm o} \pm 5^{\rm o}$.

Together with the \cii\ transition, we also detect the dust continuum of \name\ at rest-frame 160~$\mu$m. The middle panel of Figure~\ref{fig_mom} shows the integrated intensity map. The total dust continuum flux is $S_{\rm cont, 160~\mu m}=0.15\pm0.03$~mJy. The distribution of the dust emission is aligned with the morphological major axis of the \cii\ line emission, but is significantly less extended in the direction of the minor axis, although this depends on the sensitivity of the observations. A two-dimensional Gaussian fitting yields a deconvolved size (at FWHM) of $0.93\arcsec \times 0.18\arcsec$ ($\pm0.20\arcsec$) ($5.6\times1.1$~kpc) with the major axis at a PA of $27.9^{\rm o} \pm 5^{\rm o}$. In Section~\ref{Sec:cii2fir} we discuss how the differences in the spatial distribution of the dust and the \cii\ line emission can shed light on the physical conditions of the gas, and in Section~\ref{Sec:halo} we compare in more detail the light distribution from the \cii\ and dust emission.

\begin{figure*}
\centering
\includegraphics{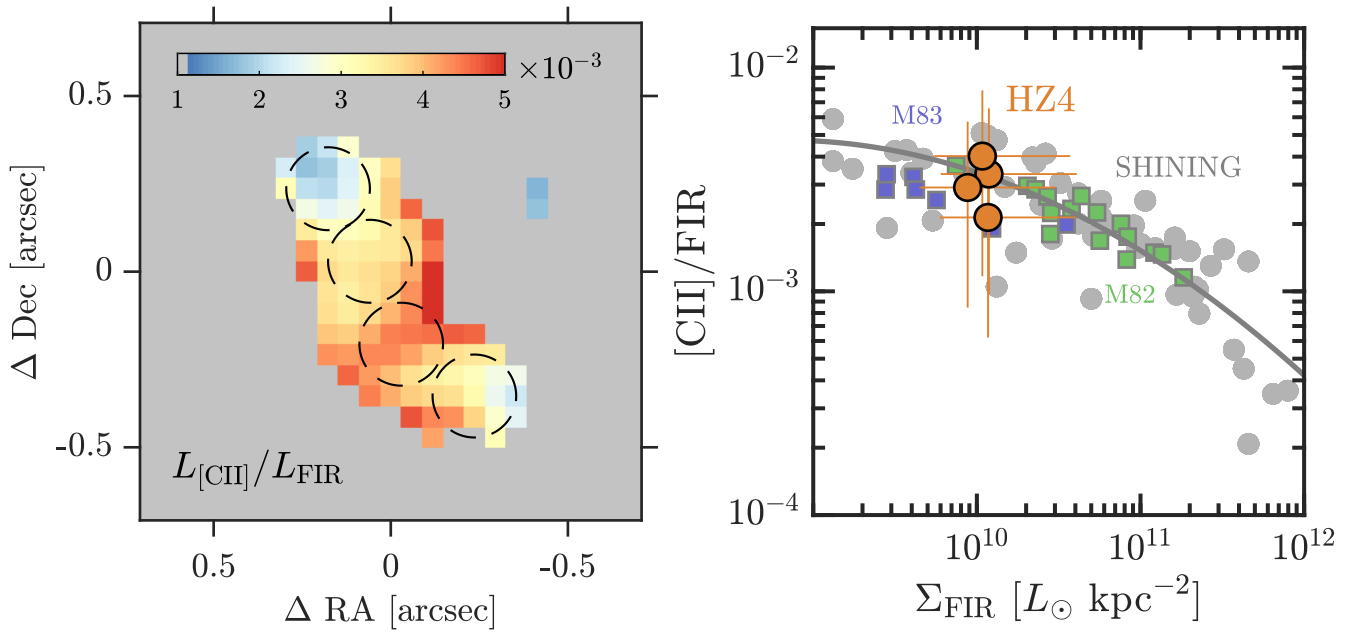}
\caption{{\it (Left)} \cii/FIR map of \name. We have considered only rest-frame $\sim160~\mu$m dust continuum emission above the 2$\sigma$ level. The colorbar indicates the \cii/FIR ratio value. {\it (Right)} \cii/FIR ratio as a function of \sfir\ in four kiloparsec-size regions extracted across the disk of \name\ (orange circles, black dashed circles in the left panel). The errorbars, mainly due to the uncertainty in the FIR luminosity, are shown in orange. For comparison, we show nearby star-forming and starburst galaxies from the SHINING sample \citep{rhc_rhc18a}, $\sim100$~pc scale regions in the central disk of M82 in green \citep{rhc_contursi13,rhc_rhc18a}, and $\sim400$~pc scale regions in the central region of M83 in purple. The gray solid line shows the best quadratic fit to the SHINING data \citep{rhc_rhc18a}.}  \label{fig_cii2fir}
\end{figure*}

In the last panel of Figure~\ref{fig_mom} we show the rest-frame UV emission observed with HST/WFC3 \citep[F160W; ][]{rhc_barisic17} plotted on top the \cii\ line integrated emission. In principle, we expect the \cii\ and UV emission to trace each other well as they correspond to important cooling and heating channels of the ISM \cite[e.g.,][]{rhc_wolfire03}. However, the original HST UV peak emission is centered at the position marked with a black dot, shifted with respect to the ALMA dust continuum and \cii\ peak emission by $\sim0.3\arcsec$. Spatial offsets of this magnitude between H$\alpha$ or UV and dust emission have been observed in $z\sim2-3$ submillimeter galaxies \citep[SMGs; e.g.,][]{rhc_hodge16,rhc_chen17}, and main-sequence galaxies at $z\gtrsim4$ \citep[e.g.,][]{rhc_maiolino15,rhc_carniani17,rhc_ginolfi20b}. \cite{rhc_fujimoto20}, for example, find an average offset between the \cii\ and UV emission of 23 ALPINE galaxies of $\sim0.15\arcsec$. The physical origin of these offsets could be related to dust obscuration or the ionization state of the gas \citep[e.g.,][]{rhc_katz17}, or stellar feedback destroying molecular clouds around star forming regions \citep{rhc_vallini15}.

Another possibility is that there are astrometric differences between the ALMA and HST datasets \citep[e.g.,][]{rhc_dunlop17,rhc_faisst20}.  According to the Cycle 6 ALMA Technical Handbook (Eq.~10.7), the astrometric precision of our ALMA observation is $\sim 0.01\arcsec$. In the case of the HST observations of \name, the astrometric precision is at the subarcsecond level \citep{rhc_barisic17}. One alternative to check the precision of the HST astrometry is to compare the position of stars in the HST field with those in the Gaia catalog \citep{rhc_gaia16,rhc_gaia18}. We find a small offset of only $\Delta {\rm RA}=-0.\arcsec07$ and $\Delta {\rm Dec}=-0.\arcsec11$, which is not sufficient to explain the $\sim0.3\arcsec$ displacement between the \cii\ and rest-frame UV disks. Therefore, the reason for at least a 0.2\arcsec ($\sim3$~kpc) spatial offset remains unclear at this point.

If we assume that the observed spatial offset between the HST and ALMA data is a problem of astrometric accuracy, we can use an alternative tracer of the stellar light to check the position of the HST rest-frame UV map. For this we use a SUBARU/HSC $i$-band image \citep{rhc_aihara19}, which has an astrometric precision better than 0.1\arcsec. To match the peak emission between the HST and SUBARU data, we need to shift the former by $\Delta {\rm RA} = -0.3\arcsec$ and $\Delta {\rm Dec} = +0.05\arcsec$. The shifted HST rest-frame UV contours are shown in the last panel of Figure~\ref{fig_mom}, and are in much better agreement with the spatial distribution of the ALMA \cii\ and dust continuum data.

Together with the dust continuum emission in \name, we detect a strong continuum source in the north-east direction at a distance of $\sim13$\arcsec\ from \name. This source has no apparent counterparts in the available HST/WFC3 images. We discuss the properties of this serendipitous source in Appendix~\ref{sec:cont_source}.

%--------------------------------------------------------------------

\subsection{ISM properties: the relationship between the \cii\ line, the FIR continuum, and the SFR in  \name} \label{Sec:cii2fir}

The \cii\ line to FIR continuum ratio is closely linked to the physical properties of the ISM in galaxies. A decrease in the \cii/FIR ratio can be driven by an increase in the ionization parameter of HII regions \citep[e.g.,][]{rhc_gracia-carpio11,rhc_rhc18b}, the impact of AGN on the ionization state of the gas \citep[e.g.,][]{rhc_langer15,rhc_rhc18b}, and/or the increase in the ratio between the radiation field intensity and the neutral gas density, which leads to the destruction or charging of small dust grains \citep[e.g.,][]{rhc_kaufman99,rhc_malhotra01,rhc_diaz-santos17}, among other factors. Typical \cii/FIR ratios range from $0.1-1\%$ in low-metallicity dwarfs and disks of star-forming galaxies \citep[e.g.,][]{rhc_cormier15,rhc_smith17,rhc_rhc18a} to $\sim0.01-0.1\%$ in nuclear regions of starburst systems and AGNs \citep[e.g.,][]{rhc_malhotra01,rhc_diaz-santos13,rhc_rhc18a}. 

Figure~\ref{fig_cii2fir} shows a map of the \cii/FIR ratio in \name. The FIR luminosity is calculated fitting a modified, single dust temperature greybody model to the rest-frame 160~$\mu$m continuum assuming a dust temperature of $T_{\rm dust}=45$~K and a dust emissivity index of $\beta=1.5$. To the best of our knowledge, these assumptions seem appropriate for a typical star-forming galaxy at $z\approx5$ \citep[e.g.,][]{rhc_pavesi16,rhc_faisst17}. 

The right panel of Figure~\ref{fig_cii2fir} shows the \cii/FIR ratio measured along the major axis of \name\ as a function of the FIR surface brightness. The error bars show how much the FIR luminosity changes if we vary our assumptions for the dust temperature and dust emissivity in the range $30\leq T_{\rm dust}~({\rm K}) \leq50$ and $1\leq \beta \leq2$, respectively. For comparison, we include the relatively tight ($\sim0.25$~dex) relation observed in nearby star-forming galaxies and starbursts from the SHINING sample \citep[][]{rhc_lutz16,rhc_rhc18a}. The four kiloparsec-scale regions in \name\ follow this relation, with \sfir\ and \cii/FIR ratios comparable to those observed in the central region of the starburst galaxy M~83 or the $\sim700$~pc disk around the nuclear region of M82 \citep[][]{rhc_contursi13,rhc_rhc18a}.

Interestingly, in the four apertures shown in Figure~\ref{fig_cii2fir} the distribution of the dust continuum emission is smooth, with variations in \sfir\ of only $\sim30\%$. The \cii\ emission, on the other hand, varies by a factor of $\sim2$ in the same area where the dust continuum is detected. In the photodissociation region (PDR) models of \cite{rhc_kaufman06}, a variation of \cii\ emission for a fixed \sfir, and thus a fixed FUV (6~eV$<h\nu<$13.6~eV) radiation field intensity ($G_{0}$), can be explained by a change in the total neutral gas density (\nh). For a fixed $G_{0}$, if two regions have neutral gas densities below the critical density for collisions with H atoms  \citep[$n_{\rm crit}\approx3\times10^3$~cm$^{-3}$;][]{rhc_goldsmith15}, then a lower \cii/FIR implies a lower neutral gas density. On the contrary, if two regions have neutral gas densities higher than the critical density, then a lower \cii/FIR ratio implies a higher neutral gas density. This is probably the case in the ISM of \name.

Unfortunately, to independently constrain $G_{0}$ and \nh, it is necessary to have observations of at least one additional PDR line (e.g., one of the \oi\ transitions). With the information available, the one relevant physical quantity we can constrain is the ratio between $G_{0}$ and \nh. This parameter is a proxy for the charging of the dust grains, which in turn sets the \cii/FIR ratio: a higher dust grain charge implies that photoelectrons have to overcome a higher Coulomb barrier, resulting in a decrease of the gas photoelectric heating efficiency \citep{rhc_tielens85}, and thus a decrease in the \cii/FIR ratio. Following the  relation between \sfir\ and $G_{0}/n_{\rm H}$ derived by \cite{rhc_diaz-santos17}, we estimate that the $G_{0}/n_{\rm H}$ ratio in \name\ is $\approx0.5$. This value is comparable to those observed in nearby, star-forming galaxies \citep[$G_{0}/n_{\rm H}\sim0.5-1$; e.g., ][]{rhc_contursi13,rhc_croxall13,rhc_hughes15}, and at least one order-of-magnitude lower than the ratios measured in ULIRGs and high-$z$ starbursts \citep[$G_{0}/n_{\rm H}\sim10-100$; e.g., ][]{rhc_gullberg15,rhc_diaz-santos17}. 

An alternative way to study the heating and cooling properties of the neutral gas in \name\ is by comparing the \cii\ line emission to the star formation activity. If the gas is in thermal balance, and \cii\ is the dominant coolant of the neutral gas, then the \cii\ luminosity is a measure of the total energy that is put into the gas by different processes, including FUV photons through the photoelectric effect on dust grains and mechanical heating. This physical connection is behind the observed relation between the \cii\ surface brightness ($\Sigma_{\rm [CII]}$) and the star formation rate surface density ($\Sigma_{\rm SFR}$) observed in the Milky Way \citep[e.g.,][]{rhc_pineda14} and nearby galaxies \citep[e.g.,][]{rhc_boselli02,rhc_delooze11,rhc_sargsyan12,rhc_delooze14,rhc_kapala15,rhc_rhc15,rhc_rhc18b}. 

\begin{figure}
\centering
\includegraphics[scale=0.5]{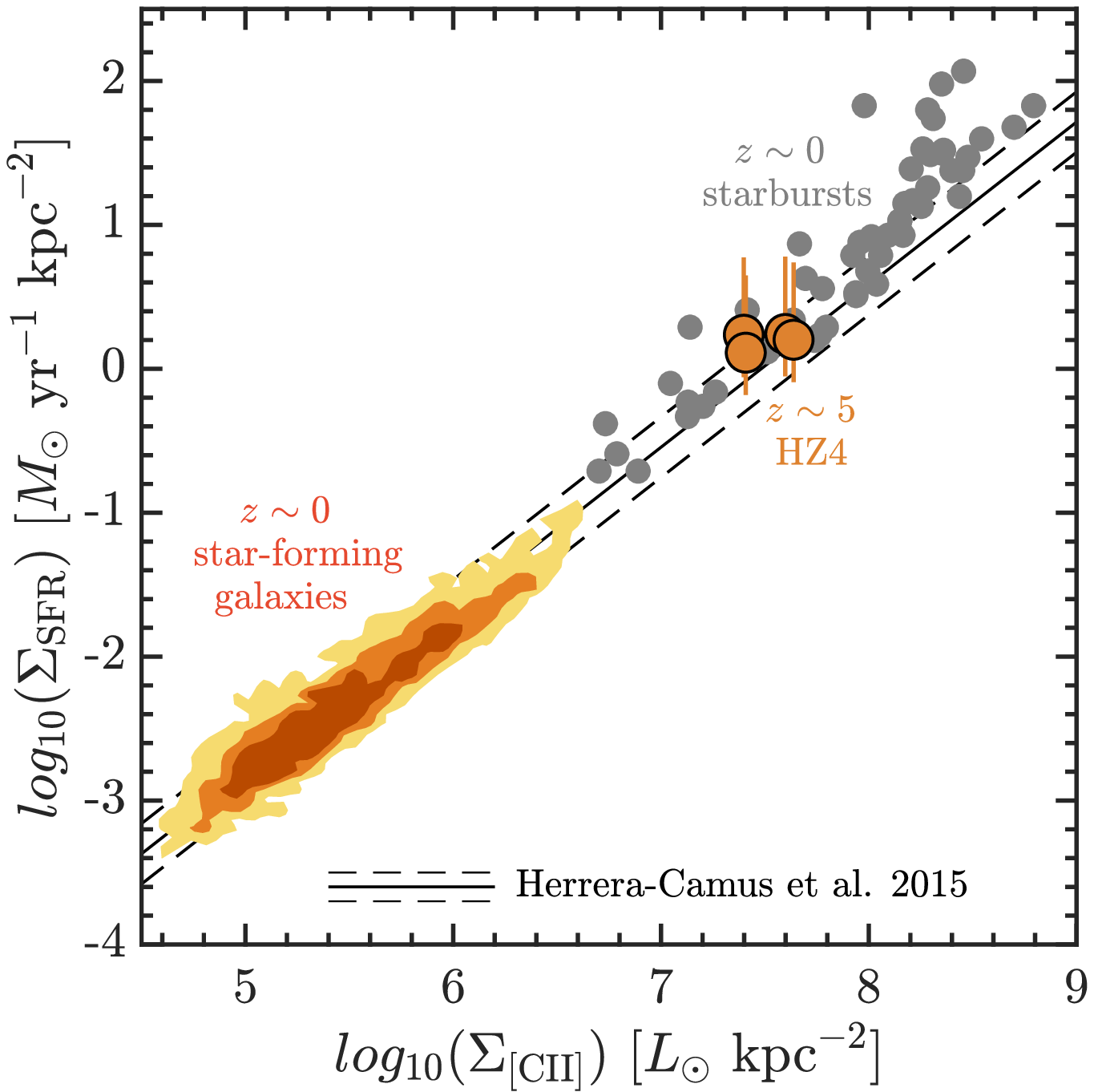}
\caption{\cii\ surface brightness density ($\Sigma_{\rm [CII]}$) versus star formation rate surface density ($\Sigma_{\rm SFR}$) observed in regions from star-forming nearby galaxies from the KINGFISH sample \citep[contours enclose 95\%, 45\%, and 25\% of the 3,846 regions; ][]{rhc_rhc15}, nearby starbursts from the SHINING sample \citep[gray circles; ][]{rhc_rhc18a}, and the four regions in \name\ where we detect the dust continuum with ALMA (orange circles). The black line shows the best fit (and $\pm1\sigma$ scatter) to the KINGFISH regions \citep{rhc_rhc15}.}\label{fig_ciisfr}
\end{figure}

Figure~\ref{fig_ciisfr} shows the $\Sigma_{\rm [CII]}-\Sigma_{\rm SFR}$ correlation observed in $\sim$kiloparsec size regions of nearby galaxies from the KINGFISH sample \citep{rhc_rhc15}, spatially resolved starbursts \citep{rhc_rhc18a}, and the four $\sim$kiloparsec-size regions of \name\ where we detect the ALMA dust continuum (see Figure~\ref{fig_cii2fir}). In the case of \name, we calculated $\Sigma_{\rm SFR}$ based on the FIR luminosity using the calibration by \cite{rhc_murphy11}, the rest-frame 160~$\mu$m luminosity following the calibration by \cite{rhc_calzetti10}, the UV luminosity using the calibration by \cite{rhc_murphy11}, and the combination of the UV and FIR luminosities following the calibration by \cite{rhc_bell05}. All four estimates of the $\Sigma_{\rm SFR}$ in \name\ are consistent within a factor of $\sim3$. The values of $\Sigma_{\rm SFR}$ shown in Figure~\ref{fig_ciisfr} correspond to those measured using the FIR luminosity.

The ratio between the FIR- and UV-based SFRs in \name, which is essentially a scaling of $L_{\rm FIR}/L_{\rm UV}$, is close to $\approx1$, as observed in other main-sequence galaxies at $z\sim5$ \citep[e.g.,][]{rhc_heinis14,rhc_capak15,rhc_bouwens16,rhc_fudamoto20}, and suggests that the observed offset between the ALMA dust continuum and HST rest-frame UV map is not caused by dust obscuration of the UV photons.

The $\Sigma_{\rm [CII]}-\Sigma_{\rm SFR}$ relation observed in nearby star-forming galaxies, characterized in Figure~\ref{fig_ciisfr} by the linear relation derived by \cite{rhc_rhc15} (Eq.~2), holds up to galaxies with $\Sigma_{\rm SFR}\sim1$~M$_{\odot}$~yr$^{-1}$~kpc$^{-2}$. This includes the central regions in \name. At higher $\Sigma_{\rm SFR}$ values, galaxies start to deviate from the local scaling relation, showing a deficit of \cii\ emission relative to $\Sigma_{\rm SFR}$. The systems that show a \cii\ deficit are typically dusty, dense starbursts characterized by strong radiation fields, high ionization parameters, and high neutral gas densities. As discussed above, all these factors play a role in driving the \cii\ deficit \citep[for a discussion see, e.g.,][]{rhc_malhotra01,rhc_gracia-carpio11,rhc_smith16,rhc_rhc18b}. 

The combination of the analysis of the \cii/FIR and \cii/SFR ratio suggests that the ISM conditions in the central region of \name\ are more comparable to those observed in modest, nearby starbursts (e.g., M82, M83, or NGC 253) rather than dense, dusty starbursts such as those found in luminous infrared systems.

\begin{figure*}
\centering
\includegraphics[scale=1.1]{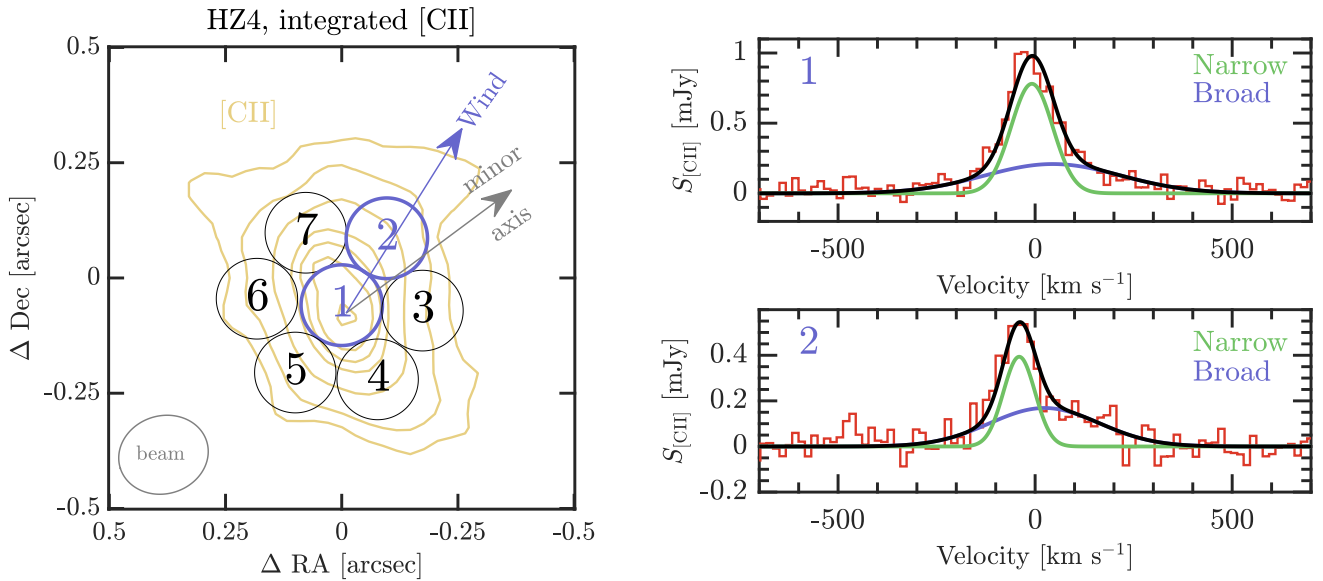}
\caption{Evidence for an outflow in \name. The left panel shows the \cii\ line integrated intensity contours with S/N$\geq3$ (yellow). The spectra extracted from the 7 apertures placed across the disk are shown in Appendix~\ref{sec:other_apertures}, except for apertures 1 and 2 (in purple), for which they are shown in the right panels. In these cases the best Gaussian fit to the spectra involves two components: a narrow component which we associate with the disk (green), and a broad component that we associate with outflowing gas (purple). The purple arrow in the \cii\ map shows the direction in the disk were we find evidence for outflowing gas, which is almost along the galaxy minor axis (gray arrow).
}  \label{fig:outflow}
\end{figure*}

% ----------------------------------------------------------------------------------------

\subsection{Outflows: evidence for an outflow along the minor axis}  \label{Sec: wind}

So far, the strongest evidence for neutral outflows in typical, star-forming galaxies at $z\sim5$ comes from the stacking of \cii\ spectra \citep{rhc_gallerani18,rhc_ginolfi20} or metal absorption lines \citep{rhc_sugahara19}. In the case of the analysis of the ALPINE galaxies, the stacked \cii\ spectra of galaxies with ${\rm SFR}>25$~M$_{\odot}$~yr$^{-1}$ show a high-velocity tail that can be fitted by a broad Gaussian component of ${\rm FWHM}\sim680$~km~s$^{-1}$ \citep{rhc_ginolfi20}. The mass outflow rate, although uncertain for reasons we discuss below, is comparable to the SFR of the stacked systems.

\begin{figure*}
\centering
\includegraphics{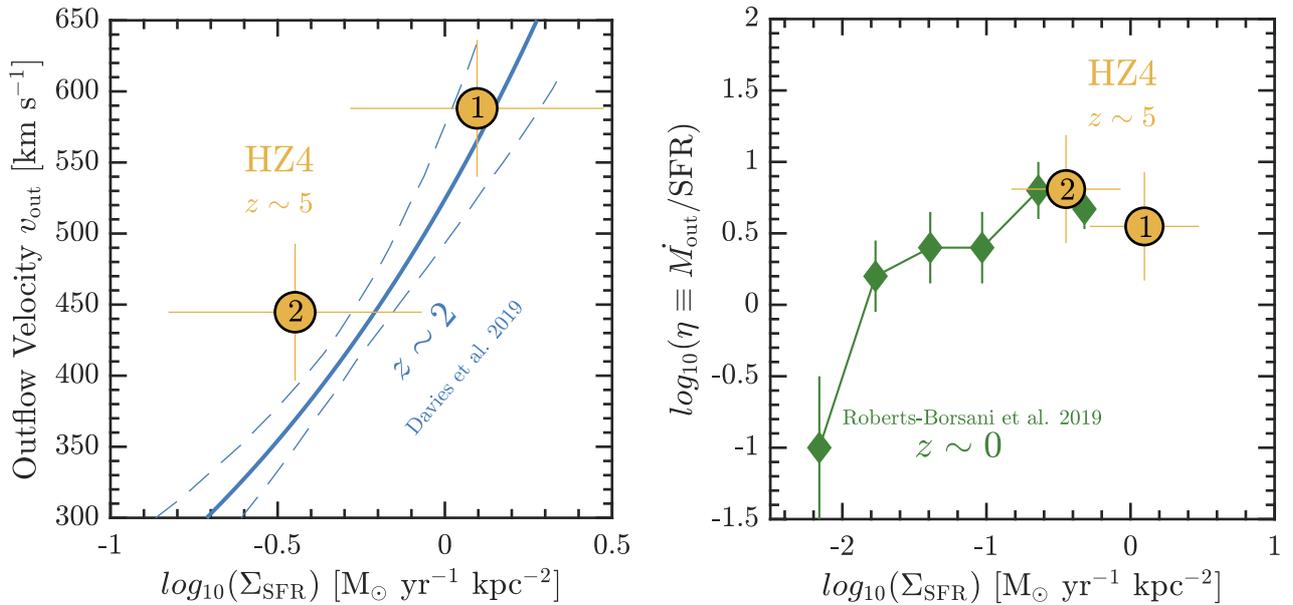}
\caption{{\it (Left)} Outflow velocity ($v_{\rm out}$) versus star formation rate surface density ($\Sigma_{\rm SFR}$) observed in apertures 1 and 2 of \name\ (orange circles) and in the stacked $\sim$kiloparsec scale star-forming regions in $z\sim2$ main-sequence galaxies \citep[blue lines; ][]{rhc_davies19}. {\it (Right)}  {\it Local} neutral mass loading factor  as a function of $\Sigma_{\rm SFR}$ for \name\ (orange circle) and stacked galaxies from the MANGA survey based on the Na~D~$\lambda\lambda5889$,5895~\AA\ absorption doublet \citep[green diamonds; ][]{rhc_roberts-borsani19}.  \label{fig:outflow_prop}}
\end{figure*}

\name\ has a total ${\rm SFR}$ of $\sim40$~M$_{\odot}$~yr$^{-1}$, comparable to the SFR in the group of stacked galaxies in \cite{rhc_ginolfi20} that show evidence for outflows. Taking advantage of the high spatial resolution and sensitivity of our observations, we search for outflow signatures on $\sim$kiloparsec scale regions across the disk of \name. Figure~\ref{fig:outflow} shows the distribution of seven circular apertures where we search for outflow signatures in the spectrum. Spectra from regions 3 to 7 (Appendix~\ref{sec:other_apertures}) look symmetric. For each spectrum from region 3 to 7, a single Gaussian fit was strongly preferred over fitting a double Gaussian distribution based on the reduced chi-squared value. We also find that the residuals from the single Gaussian fit do not show any clear evidence of flux excess over the entire velocity range. Spectra from apertures 1 and 2, on the other hand, require two Gaussian components for an optimal fit: a narrow component that can be associated with the star forming disk, and a broad component associated to a potential outflow. The right panel of Figure~\ref{fig:outflow_prop} shows the result of the double Gaussian fit. The FWHM of the broad components of apertures 1 and 2 are 406 and 326~km~s$^{-1}$, respectively. These values are a factor $\times1.6-2$ lower than the FWHM of the broad component in the stacked \cii\ spectrum of main-sequence galaxies at $z\sim4-5$ \citep{rhc_ginolfi20}, QSOs at $z\sim6$ \citep{rhc_stanley19}, and about a factor of 5 lower than the FWHM in the powerful AGN-driven outflows detected in QSOs at $z\gtrsim5$ \citep{rhc_maiolino12,rhc_bischetti19}.

It is interesting to note that the broad component features observed in Region 1 and 2 become diluted in the global \cii\ spectrum shown in Figure~\ref{fig:ciispec}. In that sense, it is only due to our ability to spatially resolve the \cii\ emission with high sensitivity that we can detect the outflow signatures. Additional evidence in favor of the outflow interpretation of the broad component comes from two signatures also observed in nearby galaxy outflows: (1) the outflow in \name\ is located at the peak of the gas velocity dispersion (see Paper II; Herrera-Camus et al. in prep.), and (2) the outflow extends nearly aligned with the minor axis \citep[the path of least resistance through the disk; e.g., ][]{rhc_chen10,rhc_leroy15,rhc_roberts-borsani20,rhc_rhc20a}. 

\begin{center}
\begin{table}[h!]
\centering
\caption{$\kappa_{\rm [CII]}(n,T,Z)$: \cii\ luminosity-to-mass conversion factor}
\begin{tabular}{c c c c} 
\hline \hline
%$\kappa_{\rm [CII]}(n,T,Z)$ 
& $T = 10^2$~K & $T = 10^3$~K & $T = 10^4$~K \\ [0.5ex] 
 \hline
 $Z=Z_{\odot}$ & & & \\
  \hline
$n = 10^2$~cm$^{-3}$ & 38.1 & 13.2 & 7.2\\
$n = 10^3$~cm$^{-3}$ & 5.8 & 2.7 & 2.0\\
$n = 10^4$~cm$^{-3}$ & 2.6 & 1.6 & 1.5\tablefootmark{a} \\
 \hline
 $Z=0.5Z_{\odot}$ & & & \\
 \hline
%& $\kappa_{\rm [CII]}(Z=Z_{\odot})$ & $= 3\times\kappa_{\rm [CII]}(Z=Z_{\odot})$ & \\
$n = 10^2$~cm$^{-3}$ & 117.5 & 40.6 & 22.16\\
$n = 10^3$~cm$^{-3}$ & 17.9 & 8.3 & 6.3\\
$n = 10^4$~cm$^{-3}$ & 7.9 & 5.0 & 4.7 \\
 \hline
 $Z=0.1Z_{\odot}$ & & & \\
 \hline
$n = 10^2$~cm$^{-3}$ & $1.0\times10^3$ & 358.9 & 195.7\\
$n = 10^3$~cm$^{-3}$ & 157.6 & 73.0 & 55.7\\
$n = 10^4$~cm$^{-3}$ & 69.6 & 44.4 & 41.7 \\ [1ex]
 \hline
\end{tabular}
%\tablefoot{The top panel shows likely members of Pismis~11. The second panel contains likely members of Alicante~5. The bottom panel displays stars outside the clusters.\\
\tablefoottext{a}{Maximal excitation case}
\label{table_kappa}
\end{table}
\end{center}

%\footnote{Maximal excitation case}

To further explore the outflow scenario, we compare the properties of the potential outflow in \name\ with those observed in the outflows of other local and high-$z$ star-forming galaxies. We first estimate the projected maximum velocity of the outflowing gas following $v_{\rm out}\sim |\Delta{v}|+{\rm FW}_{10\%}/2$ \citep[e.g.,][]{rhc_lutz20}, where $\Delta{v}$ shift of the centroid of the broad Gaussian line component for the outflow with respect to the systemic velocity, and ${\rm FW}_{10\%}$ is the full width at a tenth of the maximum of the broad outflow component. The projected outflow velocities we obtain for apertures 1 and 2 are $v_{\rm out,A1}\approx415$~km~s$^{-1}$ and $v_{\rm out,A2}\approx320$~km~s$^{-1}$, respectively. These velocities are most likely lower limits as we are not considering projection effects. Based on the axial ratio $b/a$ of the deconvolved size of the \cii\ image,  we infer an inclination for \name\ of approximately $\theta\approx45^{\rm o}$. If we assume that the outflows are perpendicular to the disk, then the deprojected velocities are a factor $1/{\rm cos{(45^{\rm o})}}\approx1.4$ larger, i.e., $v_{\rm out,A1}\approx590$~km~s$^{-1}$ and $v_{\rm out,A2}\approx445$~km~s$^{-1}$, respectively.

Next, we calculate the star formation rate surface density ($\Sigma_{\rm SFR}$). For region 1 this can be done based on the FIR luminosity or the combination of the FIR and UV continuum emission. Within the uncertainties these two are comparable. For region 2, which is not detected in the 160~$\mu$m continuum map, we estimate the SFR from the UV luminosity following the calibration by \cite{rhc_murphy11}. On average, the star formation rate surface densities of regions 1 and 2 are in the range $\Sigma_{\rm SFR}\approx0.3-1$~M$_{\odot}$~yr$^{-1}$~kpc$^{-2}$. This value is comparable to the threshold value found by \cite{rhc_newman12} for the occurrence of strong outflows in $z\sim2$ star-forming galaxies. \cite{rhc_davies19} expanded the work of \cite{rhc_newman12} by studying the ionized outflow properties of  stacked $\sim1-2$~kpc size star-forming regions in $z\sim2$ main-sequence galaxies (similar spatial scale as our ALMA observations). As Figure~\ref{fig:outflow_prop} (left panel) shows, \cite{rhc_davies19} find a positive correlation between the outflow velocities and $\Sigma_{\rm SFR}$ consistent with trends from rest-UV absorption features tracing winds at $z\sim1-3$ \citep[e.g.,][]{rhc_weiner09,rhc_kornei12}. Both regions 1 and 2 in \name\ follow this relation within the uncertainties, which provides additional support for the outflow interpretation of the broad component in the \cii\ spectra.

Another relevant quantity to study in the outflow scenario is the mass outflow rate ($\dot{M}_{\rm out}$). For this, we first need to estimate the atomic gas mass in the outflow based on the \cii\ luminosity of the broad Gaussian component. As we describe in detail in Appendix~\ref{Sec:kappa_cii}, the conversion factor \kcii\ ($M_{\rm H}=\kappa_{\rm [CII]}\times L_{\rm[CII]}$) depends on the neutral gas density, temperature, and metallicity of the gas. Figures~\ref{fig:kappa_H} and \ref{fig:Z_H} in Appendix~\ref{Sec:kappa_cii} show the variation in the value of \kcii\ as a function of the neutral gas density ($0.1\leq n_{\rm H} \leq 10^5$~cm$^{-3}$), temperature ($10 \leq T \leq 10^{4}$~K), and metallicity ($7\leq12+log_{10}({\rm O/H})\leq9$). Table~1 lists a subset of values relevant for the physical conditions that could be found in the outflowing gas of nearby and high-$z$ galaxies. In the case of maximal excitation ($T\gg91$~K, $n\gg n_{\rm crit}\sim10^{3}$~cm$^{-3}$) and gas with solar metallicity, the mass-to-light ratio is $\kappa_{\rm [CII]}\approx1.5$. This effectively represents a  lower limit to the amount of gas in the outflow traced by the \cii\ line emission \citep[see also discussion in][]{rhc_veilleux20}. If the metallicity of the gas decreases to half or one tenth solar, then $\kappa_{\rm [CII]}$ increases approximately by a factor $\times3$ or $\times27$, respectively.

\begin{figure*}
\centering
\includegraphics{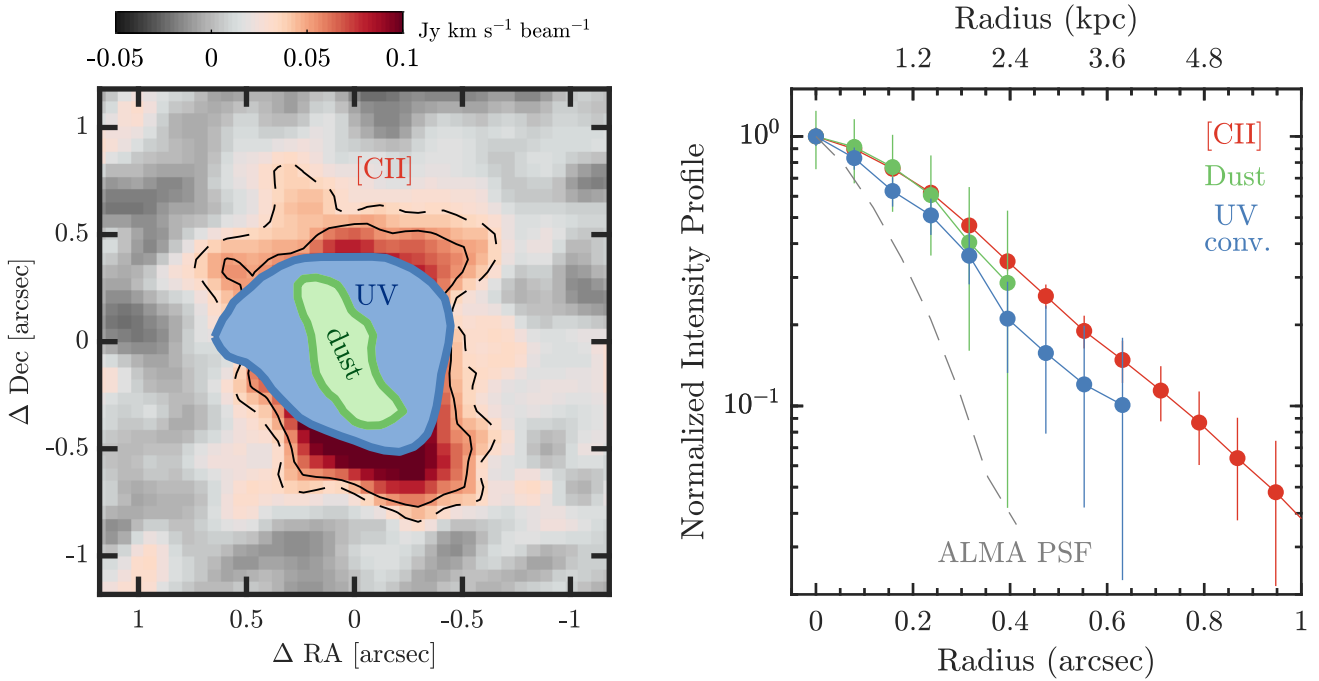}
\caption{{\it (Left)} Integrated intensity map of the \cii\ line emission. The extent of the dust and convolved rest-frame UV continuum emission with $S/N>3$ is shown in green and blue, respectively. {\it (Right)} Normalized radial intensity profile for the \cii\ line (red), dust continuum (green), and convolved rest-frame UV continuum emission (blue). The common ALMA beam intensity profile is shown with a dashed gray line. The distance from the center is shown in arcseconds (bottom) and projected kiloparsec (top, not corrected for inclination). \label{fig_sb}
}
\end{figure*}

To estimate the outflow mass in \name\ we follow a conservative approach and assume $\kappa_{\rm [CII]}\approx1.5$ (i.e., maximal excitation and solar metallicity). This results in a neutral gas mass in the outflow of $M_{\rm out,1}\approx1.2\times10^{8}$~M$_{\odot}$ and $M_{\rm out,2}\approx8.5\times10^{7}$~M$_{\odot}$ for regions 1 and 2 respectively. It could be argued based on {\it Spitzer} color-based metallicities \citep{rhc_faisst17}, or the mass-metallicity relation at $z\sim5$, that the metallicity of \name\ is about half solar or lower \citep[e.g.,][]{rhc_mannucci09,rhc_laskar11,rhc_torrey18}. This represents an additional reason to believe these mass outflow estimates correspond to lower limits.

Based on the outflow mass, we estimate the mass outflow rate as $\dot{M}_{\rm out}=M_{\rm out}\times v_{\rm out}/R_{\rm out}$. If we use the projected outflow velocity and extent of the outflow ($R_{\rm out}\sim2$~kpc), the neutral mass outflow rates for regions 1 and 2 are $\dot{M}_{\rm out,1}\approx22$~M$_{\odot}$~yr$^{-1}$ and $\dot{M}_{\rm out,2}\approx12$~M$_{\odot}$~yr$^{-1}$, respectively. If we consider inclination effects, the mass outflow rates increase by $\rm{tan(\theta)}$, so the values remain the same as ${\rm tan(45^{\rm o})}=1$. Taking into account the uncertainties and assumptions involved in the calculation, these values are comparable to the neutral mass outflow rate of $\dot{M}_{\rm out}\approx25$~M$_{\odot}$~yr$^{-1}$ measured in the stacked \cii\ spectra of the subgroup of ALPINE star-forming galaxies with median SFR similar to that of \name\ \citep[${\rm SFR}_{\rm med}\sim50$~M$_{\odot}$~yr$^{-1}$; ][]{rhc_ginolfi20}\footnote{Note that \cite{rhc_ginolfi20} assume a similar value of $\kappa_{\rm [CII]}$ for their calculations of the \cii\ outflow mass.}. This implies that the ratio between the {\it total} neutral mass outflow rate and the {\it total} star formation rate, i.e., the {\it global} neutral mass loading factor, is $\eta_{\rm global}\approx0.5$ for both \name\ and the ALPINE stack. Our present ALMA observations offer the advantage, however, to spatially resolve the regions where the outflow is located. Thus, compared to the star formation rate in regions 1 and 2, the {\it local} neutral mass outflow rates are around $\eta_{\rm local}\approx3-6$, which are a factor of $\sim6-12$ higher than the global value. As the right panel in Figure~\ref{fig:outflow_prop} shows, the {\it local} mass loading factors measured in \name\ are comparable with the neutral outflow properties of stacked MANGA $z\sim0$ galaxies with similar star formation rate surface densities \citep{rhc_roberts-borsani19}.

Compared to the modest {\it global} mass outflow rates measured in stacked $z\sim5$ galaxies, the $\sim6-10$ times larger {\it local} mass outflow rate measured in the central region of \name\ highlights the 
%The local neutral mass loading factor of $\eta_{\rm neutral}\sim5$ measured in \name\ shows the 
impact that outflows can have in locally depleting the gas, potentially preventing or delaying future episodes of star formation at their launch sites. We emphasize the word ``delay" because most likely the bulk of the gas in the outflow is not fast enough to escape the system. Following a similar approach to \cite{rhc_fluetsch19}, and based on a dynamical mass of $M_{\rm dyn}\approx10^{10.5}$~M$_{\odot}$ measured from the kinematic analysis (see Paper II; Herrera-Camus et al. in prep.), the escape velocity at a distance of one effective radius is $v_{\rm esc}\approx2\times v_{\rm out,A1}$. The fact that ejected gas remains bound to the galaxy is key for the formation and maintenance of the circumgalactic medium (CGM). Evidence for this comes directly from the existence of extended \cii\ halos around $z\sim5$ galaxies \citep[e.g.,][]{rhc_fujimoto19,rhc_fujimoto20}, which we also detect in \name. 

\subsection{Morphology: an extended component (or halo) visible in \cii\ line emission} \label{Sec:halo}

One of the most exciting new results of ALMA observations of main-sequence galaxies at $z\sim4-5$ is the detection of an extended \cii\ emission component or halo. These \cii\ halos extend significantly beyond the stellar disk and are potentially part of the circumgalactic medium \citep[e.g.,][]{rhc_fujimoto19,rhc_fujimoto20,rhc_ginolfi20}.

To compare the light distribution in \name\ of the ALMA \cii\ and dust continuum with the HST rest-frame UV continuum, we first degraded the resolution of the latter to match the resolution of the ALMA images. This was done using a kernel constructed to match the HST/WFC3 F160W point-spread function to that of the ALMA data using the IRAF task  \texttt{PSFMATCH}. The left panel of  Figure~\ref{fig_sb} shows the result, with the extent of the dust (green) and convolved UV rest-frame continuum (blue) emission with $S/N\geq3$ plotted on top the \cii\ integrated intensity map. The detection of an extended \cii\ component is evident from visual inspection.

To better quantify the distribution of the \cii, dust, and convolved rest-frame UV continuum emission we compare their azimuthally averaged radial intensity profiles calculated using the task \texttt{casairing}.\footnote{This task is provided by the Nordic ALMA regional centre: \url{https://www.oso.nordic-alma.se/software-tools.php}} The right panel of Figure~\ref{fig_sb} shows the results, with the surface brightness profile of the \cii\ line extending to a radial distance of $\sim6$~kpc, an additional $\sim3.5$~kpc than the dust continuum, and $\sim2$~kpc than the convolved rest-frame UV emission.

Fitting an exponential function of the form $e^{-r/h}$ (where $h$ is the scale-length) reproduces well the emission profiles (goodness-of-fit parameter $R^{2}\gtrsim0.95$), and yields scale lengths of $h_{\rm [CII]}=2.0$~kpc, $h_{\rm dust}=1.6$~kpc, and $h_{\rm UV}=1.7$~kpc. In the exponential disk case the effective radius is $R_{\rm eff}=1.678h$, which implies that the ratio between the \cii\ and rest-frame UV effective radius is $R_{\rm eff,[CII]}/R_{\rm eff,UV}=1.2$. This ratio increases by only 10\% if we consider the radial profile along the major axis, or if we reduce the data using a natural weighting to maximize the surface brightness sensitivity. 

%This value 
The ratio $R_{\rm eff,[CII]}/R_{\rm eff,UV}$ we measure in \name\ is lower than the median value found in the ALPINE sample of $R_{\rm eff,[CII]}/R_{\rm eff,UV}=2.3$, but comparable to the ratio of $\sim1.3$ measured in the two ALPINE galaxies with stellar masses slightly lower than \name\ \citep{rhc_fujimoto20}. The good correspondence between the \cii, dust, and rest-frame UV continuum emission in the central $R\sim3$~kpc disk suggests that the main source of heating is the photoelectric effect on dust grains. 

In \cite{rhc_fujimoto20}, ALPINE galaxies are classified as having a \cii, dust and/or UV halo if the integrated emission of the tracer is detected above a $4\sigma$ level outside the central area defined by the size of the synthesized beam \citep[typically $\sim0.8-1\arcsec$ for ALPINE galaxies; ][]{rhc_lefevre19}. \name, part of their sample, was not identified as having a \cii\ halo, possibly because of comparatively shallower \cii\ line and dust continuum observations.

The difficulty with the ``\cii-halo'' definition in \cite{rhc_fujimoto20} is that it is purely observational, and dependent on the galaxy size relative to the beam size. A more physically driven classification criteria would be to compare the extent of the \cii\ emission relative to the stellar disk as traced by the UV emission, although this would still be dependent on the sensitivity achieved in the observations of each tracer. In the case of \name, our deep observations show that the \cii\ emission is detected above a 4$\sigma$ significance both outside the stellar disk and outside a circular aperture of $1\arcsec$ in diameter. Thus, the extended  \cii\ line emission in \name\ could be considered as a ``\cii-halo'' based on both criteria.

The sources of heating of the gas in the extended \cii\ emission of high-$z$ galaxies are still in discussion \cite[for example see][]{rhc_fujimoto20}. In the case of \name, the existence of \cii\ emission beyond the UV disk could imply that UV photons are able to travel far from the star-forming regions due to the low dust-to-gas ratio of the ISM. In addition to photoelectric heating, supernova-driven cooling outflows could contribute to the \cii\ emission in the outer disk. In the one dimensional models of \cite{rhc_pizzati20}, a galaxy with a mass loading factor of $\sim3$ and a dynamical mass of $M_{\rm dyn}\approx10^{11}~M_{\odot}$ (similar to \name, see Paper~II; Herrera-Camus et al in prep.) can produce a \cii-halo with an extent comparable to that observed in \name. Finally, there could be the additional contribution to the extended \cii\ emission of shock heating from inflowing gas, although this cannot be constrained by our data.

Certainly, more observations are needed to explore the origin and physical conditions of the \cii-halo, including tracers such as the \oiii~88~$\mu$m line that could help to constrain the in-situ star formation scenario, or deeper HST or James Webb Space Telescope images that could offer a less obscured view of the stellar population or reveal reveal the presence of satellite galaxies.

\section{Summary and Conclusions}

We have obtained deep, kiloparsec scale resolution ALMA observations of the \cii~158~$\mu$m transition and the dust continuum of \name, a main-sequence galaxy at $z=5.5$ (Figures~\ref{fig:ciispec} and \ref{fig_mom}). The results are presented in two papers. This first one focuses on the properties of the ISM, outflow, and \cii-halo. The second paper will focus on the analysis of the spatially resolved kinematics of the source (Paper~II; Herrera-Camus et al. in prep.).

The main results can be summarized in three categories:

\begin{enumerate}
\item {\bf Interstellar medium properties:} The \cii\ transition is one of the major coolants of the cold neutral gas, and in combination with the FIR luminosity and/or the SFR can be used to constrain the heating-cooling balance in the ISM. Over the central $\sim$4 kpc region of \name, the \cii/FIR ratio is $\sim3\times10^{-3}$, which is comparable to the \cii/FIR ratio observed in nearby galaxies with similar FIR luminosity surface densities (Figure~\ref{fig_cii2fir}). \name\ also follows the $\Sigma_{\rm [CII]}-\Sigma_{\rm SFR}$ correlation observed in nearby star-forming galaxies and starbursts \citep[Figure~\ref{fig_ciisfr}; ][]{rhc_rhc15,rhc_rhc18b}. In this sense, \name\ does not show a ``\cii-deficit" as observed in dense starbursts, submillimeter galaxies, or a number of $z\gtrsim7$ low-metallicity systems. This suggests that the ISM properties of \name\ are comparable to those observed in the central regions of modest starbursts such as M82 or M83.

\medskip

\item {\bf Outflows:} So far	, evidence for outflows in $z\approx4-6$ normal galaxies has been presented only from stacking of \cii\ or metal absorption lines \citep[e.g., ][]{rhc_gallerani18,rhc_sugahara19,rhc_ginolfi20}. Here we present the first evidence for a star-formation driven outflow in an individual main-sequence galaxy at $z\approx5$. The outflow signatures are found in two connected regions of $\sim2$~kpc size that extend from the central part of the galaxy in the direction of the minor axis (Figure~\ref{fig:outflow}). These regions have $\Sigma_{\rm SFR}\approx0.3-1$~M$_{\odot}$~yr$^{-1}$~kpc$^{-2}$, a known threshold for the launch of star-formation driven outflows at $z\sim2$ \cite[e.g., ][]{rhc_newman12}. The projected velocity of the outflows is $v_{\rm out}\approx320-420$~km~s$^{-1}$, consistent with outflow velocities measured in $z\sim2$ main-sequence galaxies with comparable star formation activity \citep[Figure~\ref{fig:outflow_prop}; ][]{rhc_davies19}. Similar to studies based on stacking of the \cii\ spectra, we measure a modest {\it global} neutral mass-loading factor of $\eta_{\rm neutral}\sim0.5$. However, when we consider the star formation activity in the central region where the outflow is launched, the mass-loading factor increases by an order-of-magnitude. This high mass-loading factor may contribute to suppressing or delaying the star formation activity in the central region of the galaxy. 

\medskip

\item {\bf \cii\ extended emission or ``halo":} The \cii\ emission in \name\ extends beyond the dust and UV continuum disk, forming an extended component or halo of emission of about $\sim$12~kpc in diameter (Figure~\ref{fig_sb}). %This \cii-halo is similar to those detected in stacked and individual galaxies in the ALPINE survey \citep{rhc_fujimoto19,rhc_fujimoto20}. 
One of the main candidates to explain the presence of the extended \cii\ emission are the outflows detected in \name\ which have maximum velocities that are roughly half the escape velocity of the system. 

\end{enumerate}

In summary, our new ALMA observations present one of the most detailed views on kiloparsec scales of a typical galaxy when the Universe was only $\sim1$~Gyr old. We show that the combination of the \cii\ line, the dust continuum, and the rest-frame UV emission observed with HST (and in the future with the James Webb Space Telescope) offers an excellent opportunity to study different processes involved in the baryon cycle of galaxies. An important next step is to increase the number of spatially-resolved observations of main-sequence galaxies at $z\gtrsim4$, making sure to achieve enough sensitivity to detect the dust continuum and potential faint features associated to outflows and halos. 

\begin{acknowledgements}
We thank the referee for very useful comments and suggestions that improved the manuscript. R.H.-C. would also like to deeply thank his wife Fares and daughter Olivia for their support, as this paper was written at home during lockdown due to the COVID-19 pandemic. R.H.-C. would also like to thank Guido Roberts-Borsani for kindly providing the data of the MANGA galaxies. This paper makes use of the following ALMA data: ADS/JAO.ALMA\#2018.1.01605.S. ALMA is a partnership of ESO (representing its member states), NSF (USA) and NINS (Japan), together with NRC (Canada), MOST and ASIAA (Taiwan), and KASI (Republic of Korea), in cooperation with the Republic of Chile. The Joint ALMA Observatory is operated by ESO, AUI/NRAO and NAOJ. PACS has been developed by a consortium of institutes led by MPE (Germany) and including UVIE (Austria); KU Leuven, CSL, IMEC (Belgium); CEA, LAM (France); MPIA (Germany); INAF-IFSI/OAA/OAP/OAT, LENS, SISSA (Italy); IAC (Spain). This development has been supported by the funding agencies BMVIT (Austria), ESA-PRODEX (Belgium), CEA/CNES (France), DLR (Germany), ASI/INAF (Italy), and CICYT/MCYT (Spain). \end{acknowledgements}
\begin{appendix}

\section{Serendipitous detection of a strong continuum source with no HST counterpart}\label{sec:cont_source}

\begin{figure*}
\centering
\includegraphics[scale=0.9]{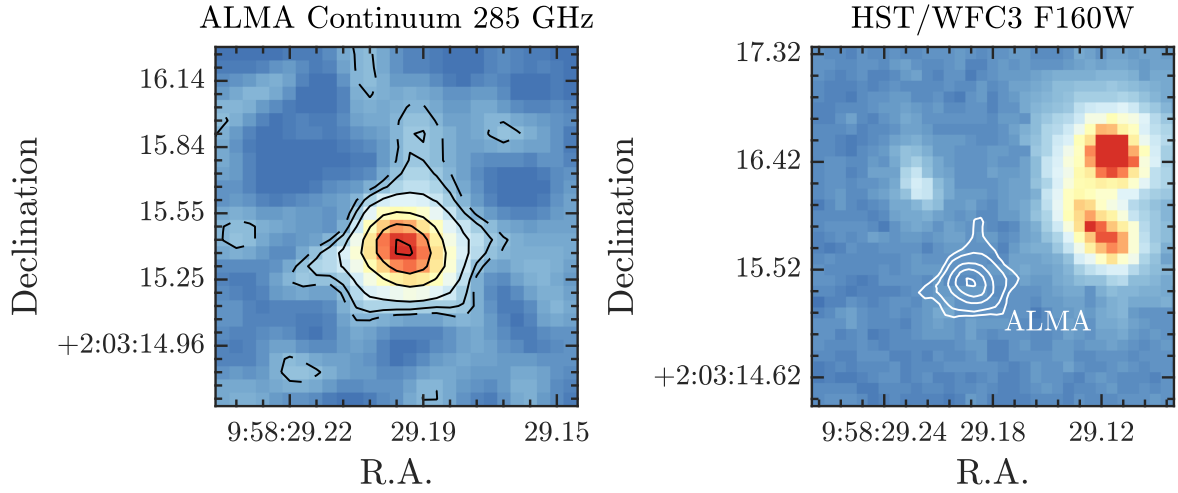}
\caption{({\it Left}) ALMA Band 7 continuum source detected at the edge of the map in the north-east direction. The contours correspond to $S/N= 2 (dashed), 3, 5, 10, 15$, and 20. ({\it Right}) Contours of the ALMA continuum source (in white) overplotted on an HST/WFC3 F160W image. There is no HST counterpart.}  \label{fig:cont_source}
\end{figure*}

While analyzing the ALMA Band 7 continuum map centered at an observed frequency of $\nu_{\rm obs}=285$~GHz, we detected a strong continuum source at the very edge of the image in the north-east direction at a distance of $\sim13$\arcsec\ at RA 09:58:29.19, Dec. +2.03.15.39. The left panel of Figure~\ref{fig:cont_source} shows the source which is detected at high significance. The source is barely resolved, and has a size (deconvolved from beam) of $0.29\arcsec \times 0.17\arcsec$. The integrated flux at 285~GHz is 0.67~mJy. We searched for a counterpart in deep HST/WFC3 images, without success (see right panel of Figure~\ref{fig:cont_source}).

\section{\cii\ spectra from apertures 3, 4, 5, 6 and 7}\label{sec:other_apertures}

\begin{figure*}
\centering
\includegraphics[scale=0.6]{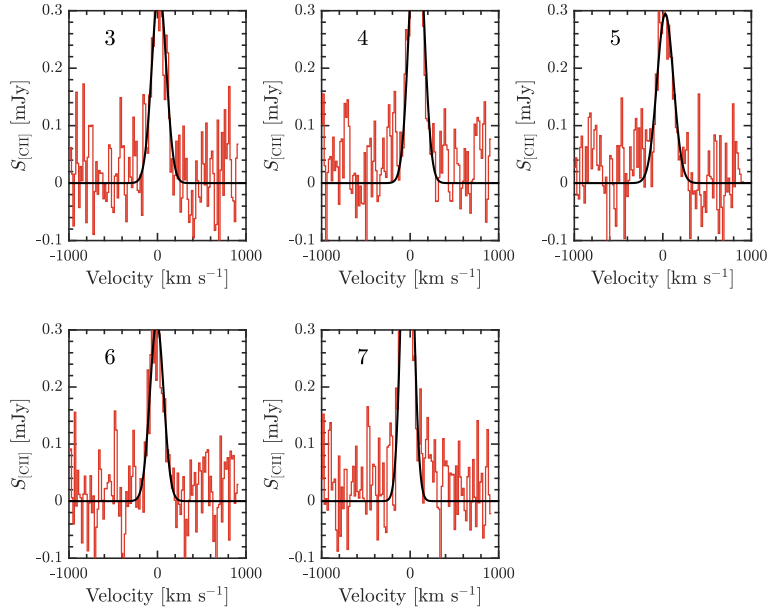}
\caption{\cii\ spectra extracted from apertures 3 to 7. The position of the apertures is shown in the left panel of Figure~\ref{fig:outflow}. The black solid lines corresponds to the best single Gaussian fit to the spectrum.}  \label{fig:other_ap}
\end{figure*}

Figure~\ref{fig:other_ap} shows \cii\ spectra extracted from apertures 3 to 7. The spatial distribution of the apertures is shown in the left panel of Figure~\ref{fig:outflow}. Different from the spectra extracted from apertures 1 and 2 which require a combination of a narrow and broad Gaussian component, these spectra are well fitted by a single Gaussian.

\section{Using the \cii\ luminosity in the outflow to estimate (a lower limit of) the outflow mass} \label{Sec:kappa_cii}

As described, for example, in \cite{rhc_crawford85}, the integrated \cii\ line intensity  from collisional excitation in the optically thin limit and negligible background emission is given by

\begin{equation}
I_{\rm [CII]} = \frac{h \nu_{ul} A_{ul}}{4\pi}\Bigg[\frac{g_{u}/g_{l} e^{-h\nu_{ul}/kT}}{1+g_{u}/g_{l} e^{-h\nu_{ul}/kT}+n_{\rm crit}(T)/n} \Bigg]N_{\rm C^{+}}.
\end{equation}
%\chi_{\rm C^{+}}N_{\rm C^{+}}

\noindent The ratio of the statistical weights of the upper and lower level is $g_{u}/g_{l}=2$, the transition energy of the \cii\ transition ($\nu_{ul}=1900.5369$~GHz) is $h\nu_{ul}=1.26\times 10^{-14}$~ergs, and the Einstein coefficient for spontaneous emission is $A_{ul}=2.36\times10^{-6}$~s$^{-1}$. $N_{\rm C^{+}}$ (cm$^{-2}$) is the column density of C$^+$, 
%and $\Phi_{\rm beam}$ is the beam area filled with C$^+$ gas. 
$n$ is the number density of the collisional partner, and $n_{\rm crit}(T)$ is the critical density defined as the ratio of $A_{ul}$ and the collision rate for de-excitation out of that state ($R_{ul}$), which is a function of the temperature of the gas \citep[e.g.,][]{rhc_goldsmith15}.
%According to \cite{rhc_goldsmith15}, for collisions with electrons and H atoms at $T=8000$~K the critical density is 44~cm$^{-3}$ and 1600~cm$^{-3}$, respectively. For collisions with electrons, H atoms, and H$_2$ molecules at $T=100$~K, the critical density is 9~cm$^{-3}$, 3000~cm$^{-3}$, and 6100~cm$^{-3}$, respectively.

Filling in the constants, and replacing $N_{\rm C^{+}}=\chi_{\rm C^{+}}N_{\rm H}$, where $\chi_{\rm C^{+}}(Z)$ is the carbon abundance (which is a function of the metallicity $Z$ of the gas) and $N_{\rm H}$ is the column density of hydrogen nuclei in the C$^{+}$ region, we obtain

\begin{equation}
I_{\rm [CII]}~({\rm erg~s^{-1}~cm^{-2}~sr^{-1}})  =2.36\times10^{-21}\Bigg[\frac{2 e^{-91~{\rm K}/T}}{1+2e^{-91~{\rm K}/T}+n_{\rm crit}(T)/n} \Bigg]\chi_{\rm C^{+}}(Z)N_{\rm H}.\label{eq:Icii_2}
\end{equation}

%The gas mass in a given projected area $A$ is $M_{\rm gas}=\mu m_{\rm H} N_{\rm H} A$, where $m_{\rm H}$ is the mass of the H nucleon and $\mu$ is the mean weight per H atom. Thus, the gas mass traced by the \cii\ line emission can be written as $M_{\rm gas}=\kappa_{\rm [CII]}\times L_{\rm [CII]}$ where the mass-to-light conversion factor $\kappa_{\rm [CII]}$ is defined as

The gas mass traced by the \cii\ line emission can be written as $M_{\rm gas}=\kappa_{\rm [CII]}\times L_{\rm [CII]}$ where the mass-to-light conversion factor $\kappa_{\rm [CII]}$ is \cite[see also][]{rhc_hailey-dunsheath10,rhc_veilleux20}:

\begin{equation}
\kappa_{\rm [CII]}~(M_{\odot}~L_{\odot}^{-1}) = 0.98\Bigg(\frac{\mu}{1.36}\Bigg)\Bigg(\frac{1.5\times10^{-4}}{\chi_{\rm C^{+}}(Z)}\Bigg)\Bigg[\frac{1+2e^{-91~{\rm K}/T}+n_{\rm crit}(T)/n}{2e^{-91~{\rm K}/T}} \Bigg].\label{eq:Mcii}
\end{equation}

\noindent \kcii\ is a function of the density of the collisional partner ($n$), the temperature (T), and metallicity ($\chi_{\rm C^{+}}(Z)$) of the gas. We have assumed a standard composition for the gas of 36\% helium ($\mu=1.36$).

We calculate \kcii\ as a function of $n, T$ and $Z$ for gas where the collisional excitation of C$^{+}$ is dominated by hydrogen atoms (H$^0$). For this we substitute $n_{\rm crit}$ using the parameterization of \cite{rhc_goldsmith15}, i.e., 

\begin{equation} \label{eq:ncrit_H}
n_{\rm crit}({\rm H^0}) = 5.9\times10^4(16+0.35T^{0.5}+48T^{-1})^{-1}~{\rm cm^{3}}.
\end{equation}

\noindent For the carbon abundance $\chi_{\rm C^{+}}(Z)$, we assume that it changes as a function of the oxygen abundance following the analytic function from the MAPPINGS photoionization code \citep{rhc_nicholls17}:

\begin{equation} \label{eq:Z}
{\rm log(C/H)}\!=\!{\rm log(O/H)}\!+\!{\rm log(10^{-1.00}+10^{(2.72+log(O/H))}}).
\end{equation}

\noindent This function has been renormalized to obtain the measured local Galactic depleted ISM carbon abundance \citep[$\chi_{\rm C^{+}}=1.5\times10^{-4}$; ][]{rhc_gerin15} when using the oxygen gas-phase abundance of the Orion nebula \citep[$12+{\rm log(O/H)}=8.65$;][]{rhc_sd11}.

\begin{figure}
\centering
\includegraphics[scale=0.5]{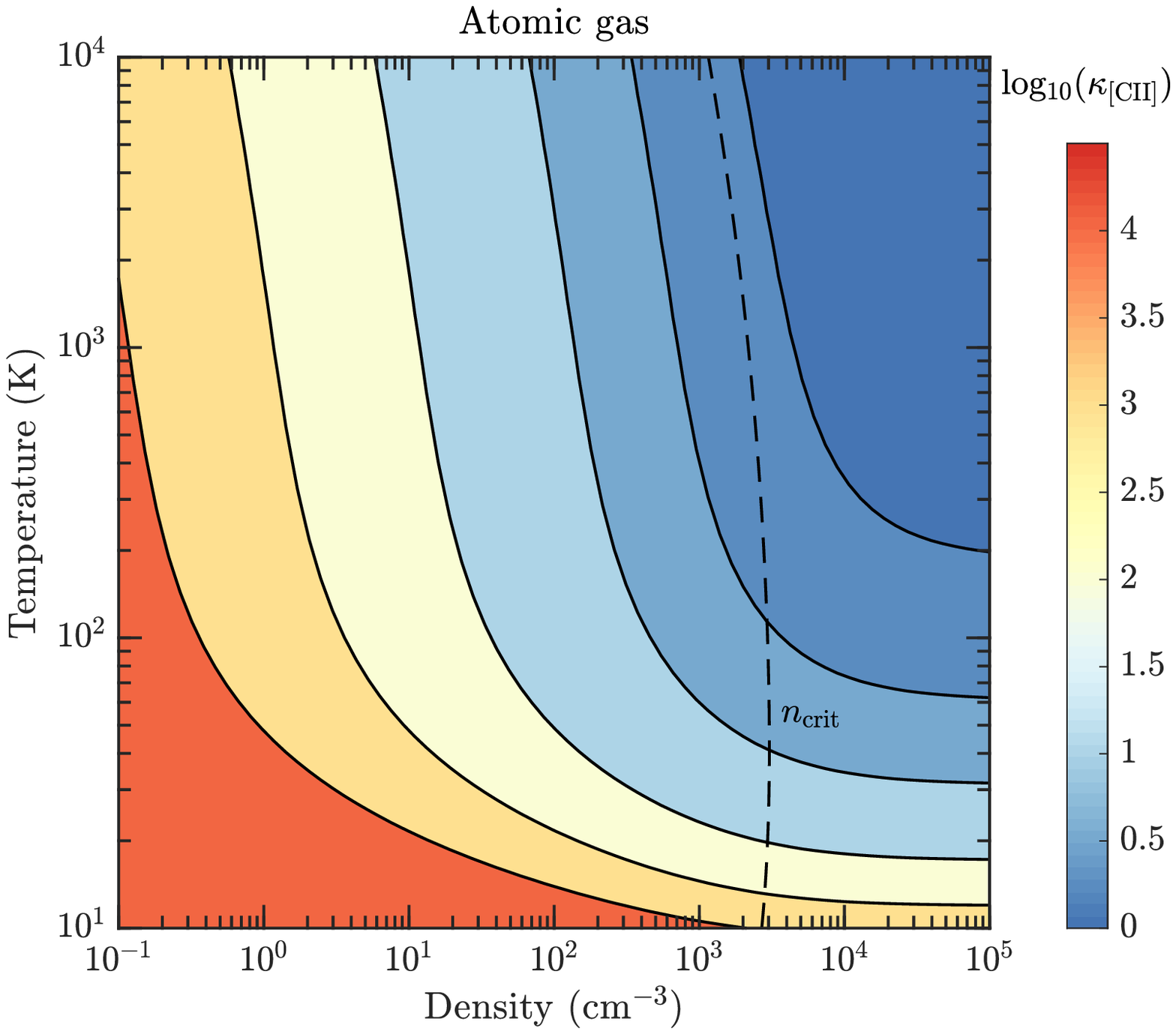}
\caption{\kcii(H$^0$) as a function of neutral gas density $n$ and temperature $T$. The value of the critical density as a function of $T$ from \cite{rhc_goldsmith15} is shown as a dashed black line. The black contours represent $log_{10}(\kappa_{\rm [CII]}/M_{\odot}~L_{\odot}^{-1})=0.25,0.5,1,2,3,4$.}  \label{fig:kappa_H}
\end{figure}

Figure~\ref{fig:kappa_H} shows the results for \kcii(H$^0$) in the parameter space $0.1\leq n~ ({\rm cm}^{-3}) \leq10^5$ and $10\leq T~({\rm K}) \leq10^4$ assuming solar metallicity. 
%. We  assume solar metallicity ($\chi_{\rm C^{+}}=1.5\times10^{-4}$) and a standard composition of 36\% helium ($\mu_{\rm atomic}=1.36$). 
The dashed black line shows the critical density as a function of temperature (Eq.~\ref{eq:ncrit_H}). \kcii(H$^0$) varies from $\kappa_{\rm [CII]}\sim1$ in dense, warm environments (maximal excitation; $n\approx10^{4-5}~{\rm cm}^{-3}, T\approx300-10^4$~K) to \kcii(H$^0)\sim10-300$ in colder gas ($T\lesssim100$~K). As expected, at gas densities $n>n_{\rm crit}$ the value of \kcii(H$^0$) only varies slightly with $n$.

\begin{figure}
\centering
\includegraphics[scale=0.5]{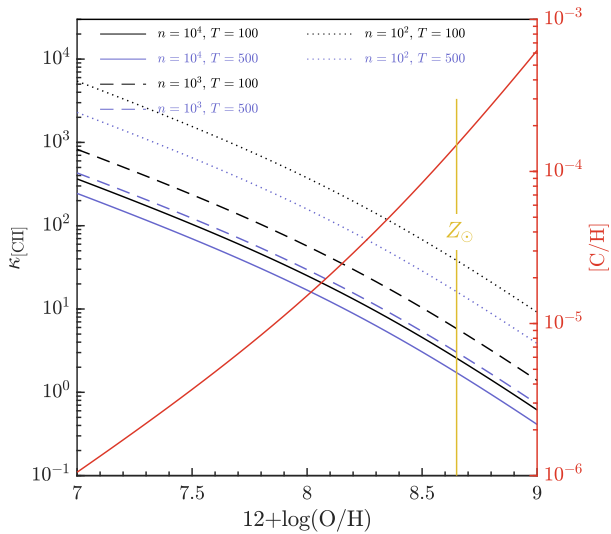}
\caption{Variation of \kcii(H$^0$) calculated for six combinations of $T$ and $n$ as a function of metallicity as traced by the oxygen abundance. The vertical yellow line shows the assumed value for solar metallicity \citep[$12+{\rm log(O/H)}=8.65$;][]{rhc_sd11}. The right ordinate shows in red the dependence on the carbon abundance [C/H] as a function of metallicity from the MAPPINGS code \citep{rhc_nicholls17}.}  \label{fig:Z_H}
\end{figure}

Figure~\ref{fig:Z_H} shows the dependence of \kcii(H$^0$) with metallicity as measured by the oxygen abundance 12+log$_{10}$(O/H). The right ordinate shows the carbon abundance dependence with metallicity from the parameterization in the MAPPINGS code. \kcii(H$^0$) increases by a factor $\sim30$ if we reduce the metallicity of the gas from solar to one tenth solar.

\end{appendix}

% Bibliography -------------------------------------------------

\bibliographystyle{aa}
\bibliography{/Users/rhc/Documents/references.bib}

\end{document}